\documentclass[journal]{IEEEtran}
\ifCLASSINFOpdf
\else
\fi

\usepackage[utf8]{inputenc}
\usepackage[english]{algorithm2e}
\usepackage{adjustbox}
\usepackage[section]{placeins}

\usepackage{float}
\usepackage{siunitx}
\begin{document}
%
% paper title
% Titles are generally capitalized except for words such as a, an, and, as,
% at, but, by, for, in, nor, of, on, or, the, to and up, which are usually
% not capitalized unless they are the first or last word of the title.
% Linebreaks \\ can be used within to get better formatting as desired.
% Do not put math or special symbols in the title.
\title{Feasibility Assessment of an Optically Powered Digital Retinal Prosthesis Architecture for Retinal Ganglion Cell Stimulation}
%
%
% author names and IEEE memberships
% note positions of commas and nonbreaking spaces ( ~ ) LaTeX will not break
% a structure at a ~ so this keeps an author's name from being broken across
% two lines.
% use \thanks{} to gain access to the first footnote area
% a separate \thanks must be used for each paragraph as LaTeX2e's \thanks
% was not built to handle multiple paragraphs
%

\author{William~Lemaire,~\IEEEmembership{Graduate Member,~IEEE,}
        Maher Benhouria,
        Konin Koua,
        Wei Tong,
        Gabriel Martin-Hardy,
        Melanie Stamp,
        Kumaravelu Ganesan,
        Louis-Philippe Gauthier,
        Marwan Besrour,
        Arman Ahnood,
        David John Garrett,
        Sébastien Roy,~\IEEEmembership{Senior Member,~IEEE,}
        Michael R. Ibbotson,
        Steven Prawer,
        Réjean Fontaine,~\IEEEmembership{Senior Member,~IEEE}% <-this % stops a space
\thanks{W. Lemaire, M. Benhouria, K. Koua, G. Martin-Hardy, L.P. Gauthier, M. Besrour, and S. Roy are with Interdisciplinary Institute for Technological Innovation (3IT), Université de Sherbrooke, Sherbrooke, Quebec, Canada e-mail: (william.lemaire@usherbrooke.ca).}% <-this % stops a space
\thanks{W. Tong is with National Vision Research Institute, Australian College of Optometry, Carlton, Victoria, Australia and M. Ibbotson is also affiliated with this institution.}% <-this % stops a space
\thanks{M. Stamp, K. Ganesan, and S. Prawer are with School of Physics, The University of Melbourne, Parkville, Victoria, Australia.}% <-this % stops a space
\thanks{A. Ahnood and D.J. Garrett are with School of Engineering, RMIT University, Melbourne, Victoria, Australia.}% <-this % stops a space
\thanks{M. Ibbotson is also with Department of Optometry and Vision Sciences, The University of Melbourne, Parkville, Victoria, Australia.}
%\thanks{Manuscript received XXXX XX, XXXX; revised XXXX XX, XXXX.This research was supported by the Australian Research Council, through Linkage Grant LP160101052, Natural Sciences and Engineering Research Council of Canada (NSERC), through Collaborative Research and Development NSERC-CRD 530093 and by CMC microsystems.}
\thanks{This research was supported by the Australian Research Council, through Linkage Grant LP160101052, Natural Sciences and Engineering Research Council of Canada (NSERC), through Collaborative Research and Development NSERC-CRD 530093 and by CMC microsystems.}

}

% note the % following the last \IEEEmembership and also \thanks - 
% these prevent an unwanted space from occurring between the last author name
% and the end of the author line. i.e., if you had this:
% 
% \author{....lastname \thanks{...} \thanks{...} }
%                     ^------------^------------^----Do not want these spaces!
%
% a space would be appended to the last name and could cause every name on that
% line to be shifted left slightly. This is one of those "LaTeX things". For
% instance, "\textbf{A} \textbf{B}" will typeset as "A B" not "AB". To get
% "AB" then you have to do: "\textbf{A}\textbf{B}"
% \thanks is no different in this regard, so shield the last } of each \thanks
% that ends a line with a % and do not let a space in before the next \thanks.
% Spaces after \IEEEmembership other than the last one are OK (and needed) as
% you are supposed to have spaces between the names. For what it is worth,
% this is a minor point as most people would not even notice if the said evil
% space somehow managed to creep in.

% The paper headers
%\markboth{IEEE Transactions on Biomedical Engineering,~Vol.~XX, No.~X, August~XXXX}%
%{Shell \MakeLowercase{\textit{et al.}}: Bare Demo of IEEEtran.cls for IEEE Journals}

\markboth{}%
{Shell \MakeLowercase{\textit{et al.}}: Bare Demo of IEEEtran.cls for IEEE Journals}
% The only time the second header will appear is for the odd numbered pages
% after the title page when using the twoside option.
% 
% *** Note that you probably will NOT want to include the author's ***
% *** name in the headers of peer review papers.                   ***
% You can use \ifCLASSOPTIONpeerreview for conditional compilation here if
% you desire.

% If you want to put a publisher's ID mark on the page you can do it like
% this:
%\IEEEpubid{0000--0000/00\$00.00~\copyright~2015 IEEE}
% Remember, if you use this you must call \IEEEpubidadjcol in the second
% column for its text to clear the IEEEpubid mark.

% use for special paper notices
%\IEEEspecialpapernotice{(Invited Paper)}

% make the title area
\maketitle

% As a general rule, do not put math, special symbols or citations
% in the abstract or keywords.
\begin{abstract}
Clinical trials previously demonstrated the notable capacity to elicit visual percepts in blind patients affected with retinal diseases by electrically stimulating the remaining neurons on the retina. However, these implants restored very limited visual acuity and required transcutaneous cables traversing the eyeball, leading to reduced reliability and complex surgery with high postoperative infection risks. To overcome the limitations imposed by cables, a retinal implant architecture in which near-infrared illumination carries both power and data through the pupil to a digital stimulation controller is presented. A high efficiency multi-junction photovoltaic cell transduces the optical power to a CMOS stimulator capable of delivering flexible interleaved sequential stimulation through a diamond microelectrode array. To demonstrate the capacity to elicit a neural response with this approach while complying with the optical irradiance limit at the pupil, fluorescence imaging with a calcium indicator is used on a degenerate rat retina. The power delivered by the laser at the permissible irradiance of 4~\si{\milli\watt\per\milli\meter\squared} at 850~\si{\nano\meter} is shown to be sufficient to both power the stimulator ASIC and elicit a response in retinal ganglion cells (RGCs), with the ability to generate of up to 35~000 pulses per second at the average stimulation threshold. This confirms the feasibility of generating a response in RGCs with an infrared-powered digital architecture capable of delivering complex sequential stimulation patterns at high repetition rates, albeit with some limitations.
\end{abstract}

% Note that keywords are not normally used for peerreview papers.

% For peer review papers, you can put extra information on the cover
% page as needed:
% \ifCLASSOPTIONpeerreview
% \begin{center} \bfseries EDICS Category: 3-BBND \end{center}
% \fi
%
% For peerreview papers, this IEEEtran command inserts a page break and
% creates the second title. It will be ignored for other modes.
\IEEEpeerreviewmaketitle

\section{Introduction}

Around 250 million people in the world are affected with moderate to severe vision impairment caused by uncorrected refractive errors, cataracts, glaucoma and degenerative retinal diseases~\cite{Bourne2010}. Among them, retinal diseases such as age-related macular degeneration and retinitis pigmentosa are particularly difficult to treat due to the complex cellular organisation of this sensory membrane. The only currently approved treatment consists in functional neurostimulation to restore visual percepts by electrically stimulating the inner retinal neurons that survive the disease. 

\begin{figure*}[!ht]
\centering
\includegraphics[width=0.8\textwidth]{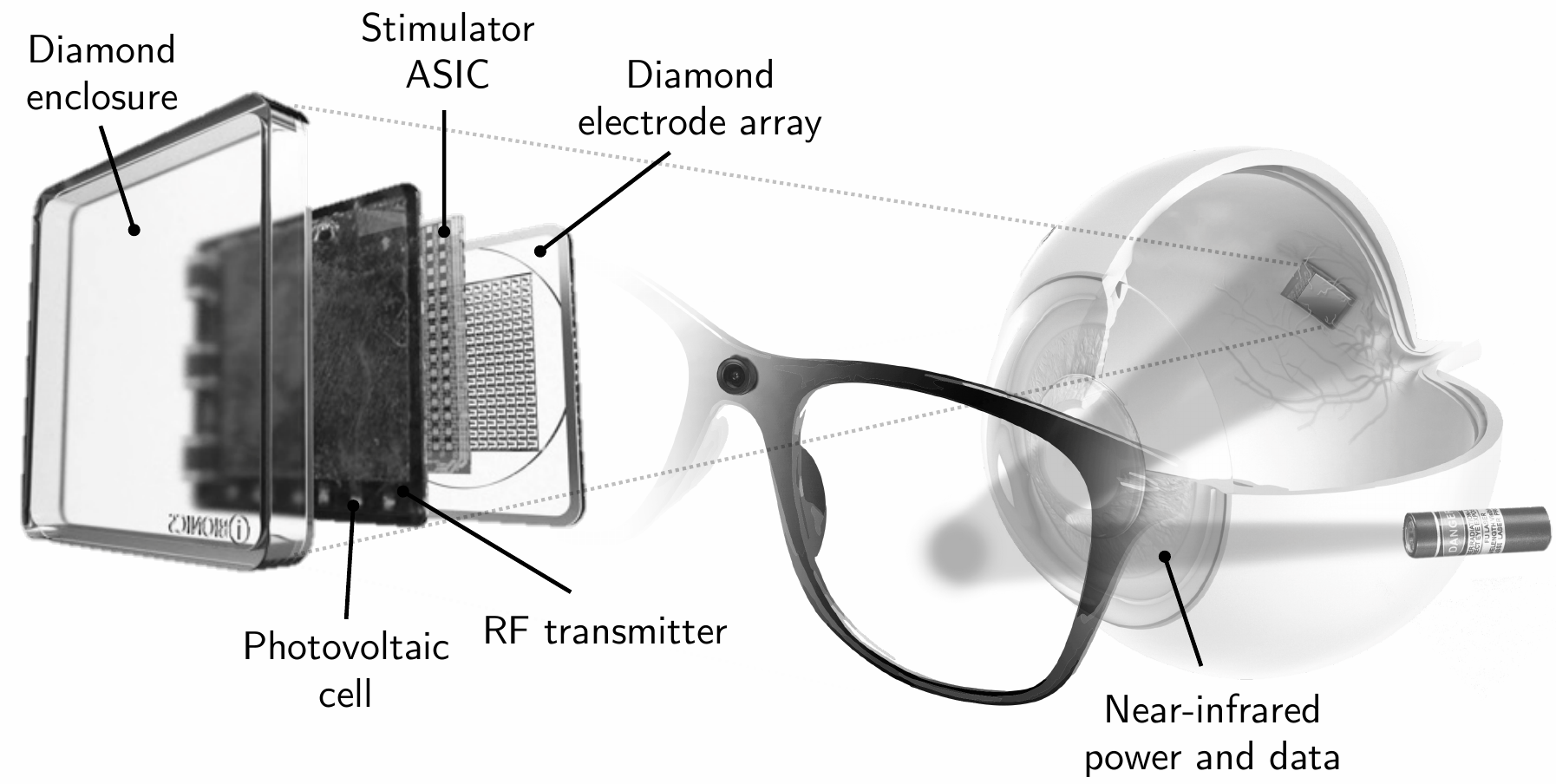}
\caption[Implant power and data delivery architecture]{Implant power and data delivery architecture. A near-infrared beam is sent through the pupil. A multi-junction photovoltaic cell captures the infrared light to power a CMOS stimulator ASIC and a photodiode recovers the data from the modulated laser beam. The ASIC delivers the stimulation through an ultrananocrystalline diamond substrate with conductive diamond electrodes.}
\label{fig:implant_overview_bw}
\end{figure*}
Existing clinically approved devices demonstrated the capacity to elicit visual percepts in patients by electrically stimulating the remaining neurons according to an image captured by a camera. They either use an external camera~\cite{DaCruz2013} (ARGUS II, Second Sight Inc., Sylmar, California, USA) or an internal photodiode array~\cite{Stingl2015} (Alpha IMS, Retina Implant AG, Reutlingen, Germany) and replicate the image with stimulation pulses on an electrode array surgically affixed to the retina. In both cases, these implants receive power through cables traversing the eyeball. While they enable the generation of visual percepts with neurostimulation, the transcutaneous cables require intricate surgery involving possible complications such as conjunctival erosion, conjunctival dehiscence (reopening of the surgical incision), hypotony (reduction of intraocular pressure) or endophthalmitis (infection) due to the permanent skin penetration~\cite{Humayun2012, Ho2015}. Moreover, the cables can lead to premature failing of the device. In the case of the alpha IMS prosthesis, the median lifetime of the cables was assessed at 1.2 years over 30 implanted first generation devices and at 7.0 years for the second generation~\cite{Daschner2017}.

To overcome the safety and reliability limitations induced by the transcutaneous cables, a wireless subretinal implant based on a microphotodiode array was previously proposed~\cite{Mathieson2012}. A camera mounted on a pair of glasses captures an image of the visual field and projects it on the retina at high intensity using an infrared projector. The photodiodes directly transduce the infrared image to stimulation pulses on electrodes to which they are individually coupled~\cite{Boinagrov2016}. The duration and intensity of the projection determine the stimulation pulse widths and currents. While effective for indiscriminate bipolar cells stimulation, this architecture might be less suitable for potential epiretinal stimulation devices aimed at delivering cell-type specific stimulation~\cite{shah_computational_2020}. In order to produce a stimulation pulse on a particular electrode, the eyeglasses would need to locate the implant with an accuracy finer than the electrode size at a high refresh rate~\cite{Palanker2005b}, which can be highly challenging considering the rapid eye saccades and frequent eyeglasses displacements~\cite{Paraskevoudi2019, Asher2007}. Although the absence of digital electronics simplifies the design of the implant and minimizes the power consumption, it limits the possibility of dynamically varying some stimulation parameters such as the interphase gap and pulse polarity for selective cell-type stimulation~\cite{Jensen2006a}. It also limits the use of active charge balancing~\cite{Greenwald2017, Sooksood2010} and the delivery of flexible multipolar stimulation patterns such as current steering~\cite{Jepson2014b, Matteucci2013} and current focusing~\cite{Fan2019}, which are demonstrated means of enhancing spatial resolution.

In order to provide wireless operation while retaining the flexibility of an implanted digital stimulation controller, we propose an implant architecture that A) receives both power and data through an optical link and B) decouples this link from the stimulation by embedding a digital controller capable of spatially confined stimulation strategies. To validate the feasibility of this power and data delivery method, a 288 electrode application-specific integrated circuit (ASIC) was designed in TSMC~CMOS~65~nm~LP~\cite{lemaire_retinal_2021} and packaged with a multijunction photovoltaic cell for power recovery. Calcium imaging fluorescence microscopy is used to validate that the device can elicit a response on retinal ganglion cells of rats affected by inherited retinal degeneration. Section II presents the implant architecture. Section III presents the materials and methods used to validate the retinal ganglion cells' (RGCs) response. Section IV presents the stimulation results and Section V discusses the implications for future implant design. 

\section{Implant Architecture}

The implant comprises multiple heterogenous components to allow photovoltaic operation (Figure \ref{fig:implant_overview_bw}). A high efficiency multi-junction photovoltaic cell recovers the optical power, and a photodiode, with a higher frequency response, receives the data transmitted by modulating the infrared beam. A stimulator ASIC then decodes the stimulation data, and executes the stimulation pattern on a 288 diamond electrode array. An embedded analog-to-digital converter (ADC) characterizes the electrode properties and sends them back to a radio-frequency (RF) receiver mounted on a pair of smart glasses through a custom-designed RF transmitter. The photovoltaic cell, photodiode, RF transmitter and passive components are assembled on a printed circuit board interposer (Figure \ref{fig:Implant_picture}), which is then mounted on the subassembly comprising the diamond array and the stimulator ASIC~(Figure \ref{fig:Implant_picture}). The next section details the rationale behind the design and the choice of each component.

\subsection{Photovoltaic Cell}

Since the retina is sensitive to temperature increases, the implant power supply is limited by the permissible optical power density that can enter the eye. Thermal damage can occur because of protein denaturation following light absorption in the retinal pigment epithelium. For an 850~nm beam entering the natural or dilated pupil, safety standards for opthalmic devices dictate that the maximum permissible radiant power is limited to $6.93\times10^{-5} C_T C_E P^{-1}$ for chronic exposure at 850~nm, where the wavelength parameter $C_T=2$ at 850~nm~\cite{Mathieson2012, Delori2007, LaserInstituteofAmerica2007}. The pupil factor $P$ models its contraction and dilatation and is equal to one at 850~nm. For spot sizes larger than 1.7 mm in diameter, $C_E=29.38$~\si{\watt\per\milli\meter\squared}. This results in a maximum permissible radiant power density of 4.06\si{\milli\watt\per\milli\meter\squared} that can enter the pupil.

Maximizing the power reaching the implant requires a high efficiency PV cell. Recent photovoltaic cells based on vertical epitaxial heterostructures achieve efficiencies up to 65~\% for monochromatic sources~\cite{Fafard2016}. By stacking multiple thin GaAs photovoltaic junctions with submicron absorption thicknesses, it is possible to achieve sufficient voltage for stimulation. The implant is designed around a 3~$\times$~3~\si{\milli\meter\squared} photovoltaic cell, resulting in a maximum usable power of 36.5~mW, given the power density limit above. Since redesigning a cell with these custom dimensions requires costly developments, a commercial bare die optical transceiver (Broadcom AFBR-POCXX4L) with dimensions of 1.7~$\times$~1.7~\si{\milli\meter\squared} was instead chosen to demonstrate the proposed architecture.

A 15~\si{\micro\farad} capacitor (C1 in Figure \ref{fig:ASIC_architecture}) stabilizes the voltage output of the photovoltaic cell and acts as an energy reservoir to complete a stimulation pulse in the event of a power loss during, for example, blinking. The photovoltaic cell connects to the ASIC (Figure \ref{fig:ASIC_architecture}) through  diode D1 (BAS116LP3-7, Diodes Incorporated) to prevent capacitor C1 from discharging into the photovoltaic cell when the laser does not reach the implant, and to prevent the PV cell maximum output of 4.4~V from exceeding the maximum supply voltage of the 65~nm technology.

\subsection{Photodiode}
In retinal prostheses, wireless data transmission is typically done with an inductive link~\cite{Mashhadi2019,Stingl2015,Chen2010}. However, the bandwidth is generally limited to hundreds of kbit/s and requires a percutaneous cable coupled with a large receiving coil. On the other hand, free-space optical communication can accommodate high data rates with a receiver of minimal complexity and size. The proposed receiving circuit is based on a transimpedance amplifier coupled to a comparator~\cite{lemaire_retinal_2021} that decodes the data from the photodiode (Albis PDCA04-100-GS). To prevent power variations during transmission and facilitate decoding, the glasses transmit the stimulation scheme using a DC-balanced Manchester code at 2 Mbits/s. The Manchester line code provides a transition in the middle of every bit interval, thus making bit clock recovery trivial at the receiver.

\subsection{Stimulator ASIC}

\begin{figure}[!ht]
\centering
\includegraphics[width=0.5\textwidth]{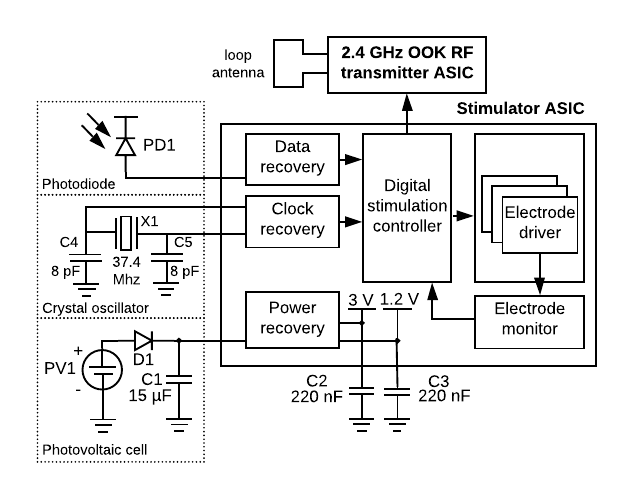}
\caption[Photovoltaic implant system schematic]{The photovoltaic cell connects to the power recovery block to capacitor C1 to ensure stability. The power recovery module linearly regulates the voltage to 3~V for the electrode drivers and 1.2~V for digital circuits. The clock recovery circuits provides a 935 kHz clock to the digital stimulation controller and a 37.4~MHz clock to the data recovery to oversample the Manchester-encoded data. From this recovered data, the digital stimulator ASIC sequences the pulse train for the electrode driver. The electrode monitor records the voltage at the output of any electrode driver and sends it out through the custom 2.4~GHz RF transmitter.}
\label{fig:ASIC_architecture}
\end{figure}

The stimulator ASIC is designed in 65~nm CMOS to allow integration of high-density digital circuits. Details about the ASIC are presented in a separate paper~\cite{lemaire_retinal_2021}. Its architecture (Figure \ref{fig:ASIC_architecture}) includes 1) 288 electrode drivers, 2) a digital stimulation controller, 3) an optical data recovery circuit, 4) a power management module and 5) an electrode characterization circuit. 

\subsubsection{Electrode driver}  To prevent irreversible electrochemical reactions at the electrode-tissue interface, the electrode driver must provide charged-balanced biphasic pulses. However, in a typical CMOS current source and sink pair, the process variations will unbalance the cathodic and anodic currents. To prevent this, the ASIC uses a dynamic current copy architecture. It operates with a calibration phase where the current sink driver sets the current that flows through the current source driver. The current source driver then copies that current and stores the calibration, corresponding to the gate-source voltage of the transistor, on a capacitor.~\cite{Chun2014, Tran2014}. The electrode driver can provide pulse widths ranging from 10~\si{\micro\second} to 700~\si{\milli\second} in steps of 10 \si{\micro\second} and with amplitudes from 50~\si{\micro\ampere} to 255~\si{\micro\ampere} in steps of 1~\si{\micro\ampere} with a voltage range of up to $\pm$ 2.7~V. 

\subsubsection{Digital stimulation controller} One of the key requirements for the stimulator ASIC is to provide flexible stimulation patterns. Moreover, because the optical power delivery can be interrupted by an eye blink, the implant must also be able to restore stimulation quickly after power up. Some implantable ASICs require a configuration phase and a stimulation phase~\cite{Tran2014}, and in the event of a power failure, this implies that the device must be reprogrammed before stimulation can resume. The digital stimulation controller operates in a stateless fashion, where each new frame fully configures the next stimulation pulses (phase durations, currents, and selection of active and return electrodes). Thus, as soon as the power is reestablished, the stimulation resumes its operation without the need for bidirectional communication.

\subsubsection{Electrode monitor} The characterization of electrode impedance enables adaptation of the stimulation to the available voltage dynamic range. To achieve this, any given electrode can be selected via a multiplexer for connection to a 8-bit ADC. To allow the waveform measurement of short pulses on the order of tens of \si{\micro\second}, it digitizes the voltage of the stimulation pulse at a maximum sampling rate of 90~kHz. 

\subsubsection{Power, data and clock recovery} The power recovery block linearly regulates the PV cell power to 3.0~V for the electrode driver and electrode monitor circuits and to 1.2~V for the digital circuits. Having two different voltages allows greater stimulation headroom while minimizing the power consumption of digital circuits. 
The clock recovery circuit generates the clock from the 37.4 MHz crystal, and divides it by 40 to provide a 935~kHz system clock. The data recovery circuit uses a transimpedance amplifier to recover the Manchester-encoded data from the photodiode, and oversamples it with the 37.4~MHz clock. Oversampling enables maximum energy transfer from the received bit and straightforward bit clock recovery (no phase-locked loop) to minimize power consumption.

\begin{figure}[!ht]
\centering
\includegraphics[width=0.45\textwidth]{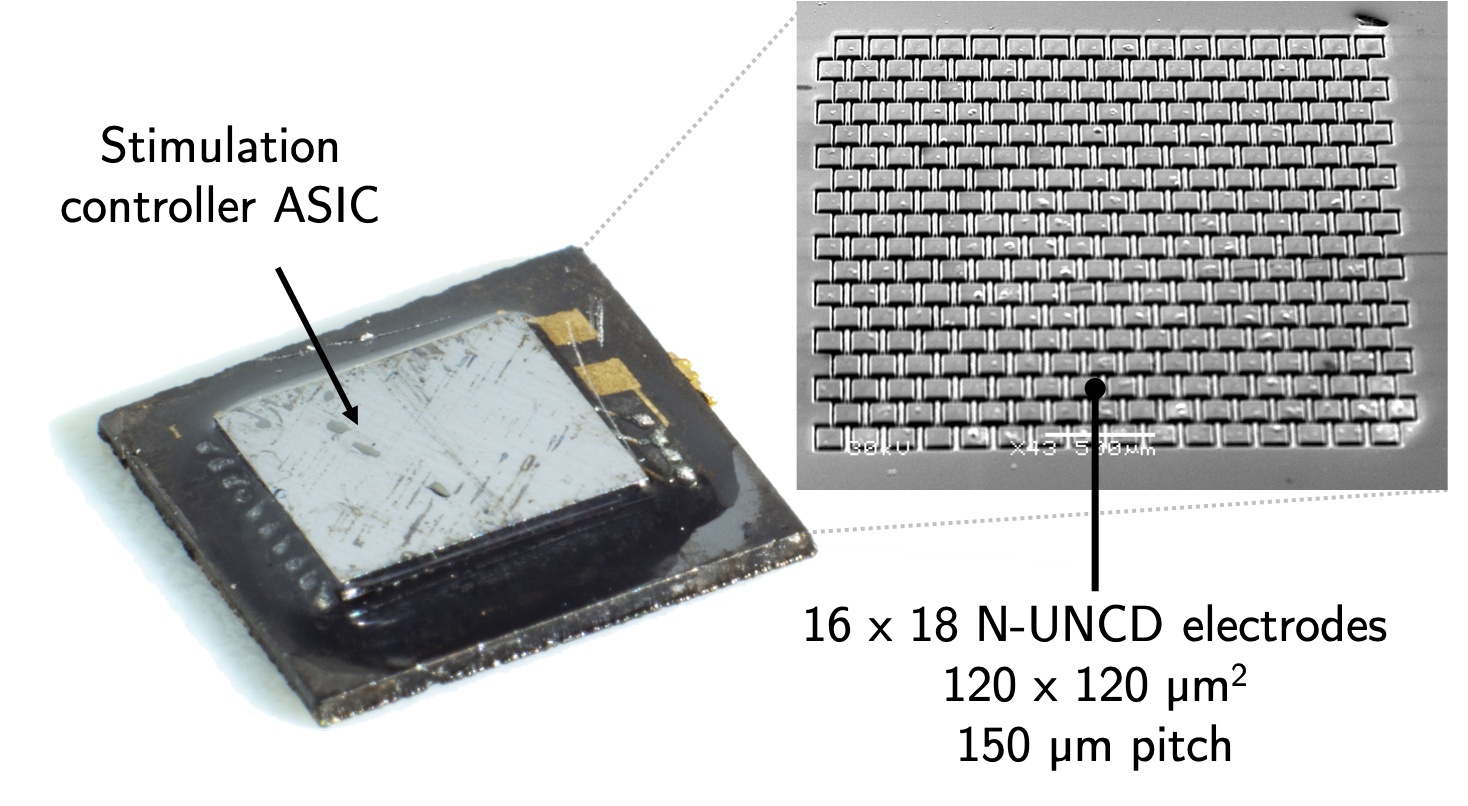}
\caption[Conceptual representation of the lower layer of the implant]{Conceptual representation of the lower layer of the implant. The stimulator ASIC is assembled on the diamond substrate with solder bumps to connect to each of the 288 electrodes. The ASIC-diamond substrate assembly process is still under development.  }
\label{fig:Diamond_and_ASIC}
\end{figure}

\begin{figure}[!ht]
\centering
\includegraphics[width=0.45\textwidth]{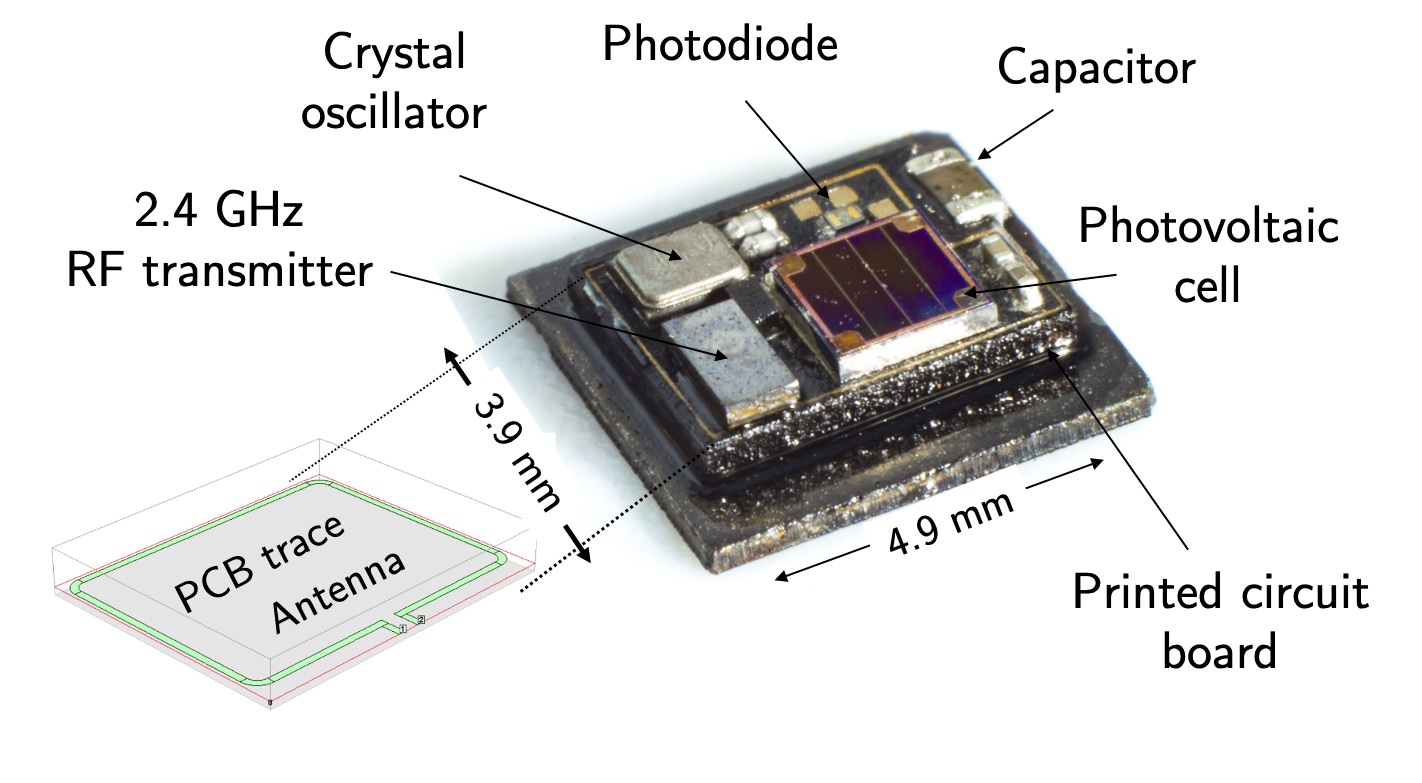}
\caption[Conceptual representation of the implant stack]{Conceptual representation of the implant stack. The photovoltaic cell, photodiode, crystal oscillator and RF transmitter are assembled on a 2-layer FR4 printed circuit board. A copper trace antenna surrounds the components. The printed circuit board is assembled on the diamond substrate (Figure \ref{fig:Diamond_and_ASIC}).}
\label{fig:Implant_picture}
\end{figure}

\begin{figure*}[!htb]
\centering
\includegraphics[width=1.0\textwidth]{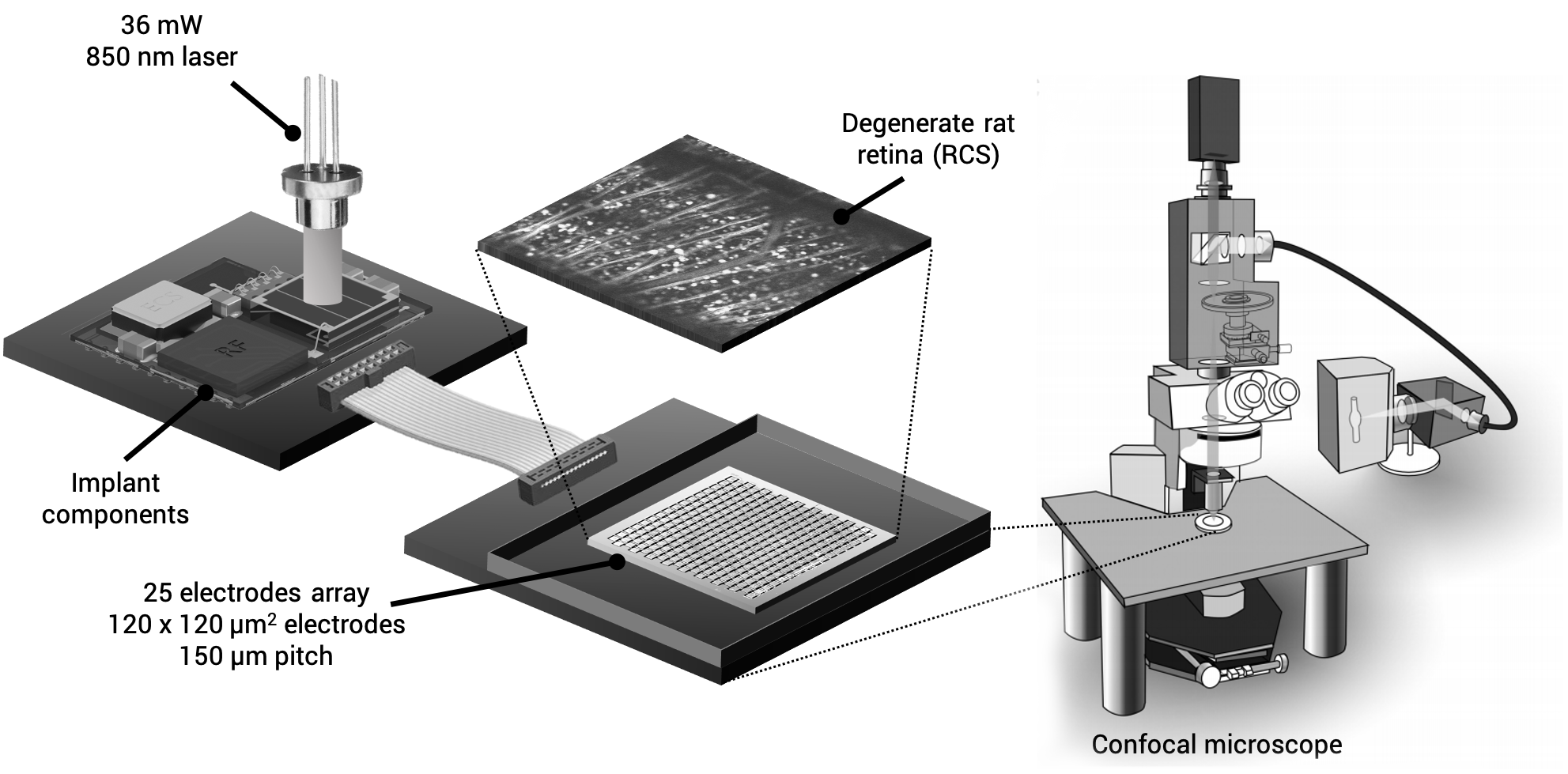}
\caption[Calcium imaging experiment configuration]{To validate the implant powering method using laser illumination, an apparatus was designed to facilitate calcium imaging where the implant components are assembled on a printed circuit board. A 35~mW, 850~nm laser powers the implant. A cable connects the implant to a 5~$\times$~5 electrode array. A degenerate rat retina stained with a calcium indicator is placed on the electrode array with retinal ganglion cells facing up. The RGCs's response is evaluated by measuring rapid fluorescence variations with a confocal microscope.}
\label{fig:methods2}
\end{figure*}

\subsection{Diamond Electrode Array and Package}

The packaging and electrode design of retinal implants is critical to ensure reliability while immersed in a biological fluid environment. The Argus II implant was enclosed in a fully hermetic package, with one cable connection to each of the 60 platinum-gray electrodes. Within three years of operation, 29 out of 30 implants were still functioning~\cite{Ho2015}. However, an implant with a significantly higher electrode count would require an excessive amount of feedthroughs with this approach. Instead, the Alpha IMS device generates stimulation waveforms directly on the pads of its CMOS chip, which are coated with iridium oxide (IrOx). Since this approach precludes the use of a hermetic enclosure, the device is instead encased in conformal coating to minimize corrosion. Without a hermetic enclosure, the median lifetime of the CMOS chip was assessed at 1.22~years~\cite{Daschner2017}.

Another possible failure mode is the electrode material degradation. Iridium Oxide and platinum electrodes are often used due to their adequate charge injection capacity and impedance for retinal stimulation. However, these materials are deposited as a coating and can be subject to delamination~\cite{Slavcheva2004}. Alternatively, ultrananocrystalline~(UNCD) diamond can be made conductive with the co-deposition of a dopant (boron) and the inclusion of nitrogen during its production by chemical vapor deposition (CVD). This electrode material provides sufficient charge injection capacity for stimulation and while allowing non-conductive and conductive diamond to coexist in the creation of a monolithic package comprising both the enclosure and the electrodes~\cite{Ahnood2017, Ganesan2014, Garrett2012}. 

Using this method, a 16~$\times$~18~diamond electrode array was designed with 120~$\times$~120~\si{\micro\meter} square electrodes separated by a pitch of 150~\si{\micro\meter} on which the stimulator ASIC will be assembled. Due to ongoing development in the ASIC-diamond substrate assembly process, the stimulator ASIC and components were instead mounted on a printed circuit board and linked via wires to a 5~$\times$~5 electrode diamond array with identical pitch and electrode dimensions. The fabrication of the diamond array is presented in a separate paper~\cite{Tong2019}.

\subsection{Printed Circuit Board Interposer}

In the complete implant, the photovoltaic cell, crystal oscillator, PV cell and RF transmitter will be assembled on a high density printed circuit board (Figure \ref{fig:Implant_picture}). The FR-4 printed circuit board (PCB) comprises 4 layers, with dimensions of 3.9~$\times$~4.9~\si{\milli\meter\squared} and a thickness of 1.6~mm. A copper trace surrounds the PCB and forms the RF antenna. This PCB will then be affixed to the diamond substrate (Figure \ref{fig:Diamond_and_ASIC}). For the feasibility study with calcium imaging experiment, the implant was physically separated from the diamond substrate and connected with a cable.

\subsection{RF Transmitter and Antenna}

\begin{figure}[h!t]
\centering
\includegraphics[width=0.25\textwidth]{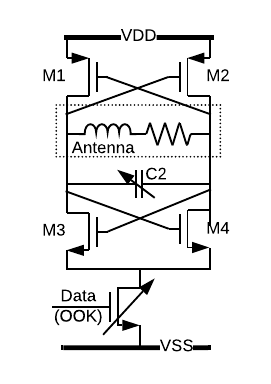}
\caption[RF transmitter circuit]{Complimentary cross-coupled LC oscillator architecture of the RF transmitter. The PCB loop antenna is modeled by a resistor and an inductor.}
\label{fig:RF_schematic}
\end{figure}

Due to power and area limitations, it is necessary to minimize the complexity of the implanted RF transmitter and antenna and relocate the complexity at the receiver side where there are less constraints. A typical oscillator-based transmitter requires multiple internal RF submodules and external components. To minimize the complexity, the transmitter operates from a simpler complementary cross-coupled LC oscillator architecture at 2.4 GHz (Fig. \ref{fig:RF_schematic}). An internal on-chip capacitor and a loop PCB antenna inductor compose the LC resonant network. Since the resonant frequency changes with fabrication variations, the on-chip capacitor is digitally tunable to adjust the frequency. The transmission efficiency at higher frequencies of 2.4~GHz allows a good compromise between tissue losses and loop antenna efficiency, although the efficiency is expected to be significantly lower in a biological environment than in air~\cite{Mercier2014}. 

The transmitter supports both on-off keying (OOK) and frequency-shift keying (FSK) modulations. In OOK mode, transistor M5 switches the oscillator according to the serial data stream. In FSK mode, the oscillator is maintained active and the frequency is modulated using tuning control signals for the on-chip capacitor. Transmission power is adjustable by gating the width of M5 to control the current passing through the cross-coupled pair. The transmitter is implemented in 65~nm~GP technology with a die size of 0.7~$\times$~1.5~\si{\milli\meter\squared} and an active area of 30~$\times$~60~\si{\micro\meter\squared}. The power consumption varies from 0.2 mW to 0.5 mW during transmission depending on the selected transmission power. 

Antenna design for implantable transmitters generally involves a compromise between the transmission efficiency and dimensions. Due to the surgery constraint, the physical antenna size must be much smaller than its radiating wavelength at 2.4 GHz. With side dimensions of a few millimeters, the loop antenna can be modeled as an inductor in series with a resistor. The antenna dimensions and materials allow to estimate its characteristics. The antenna was fabricated with dimensions of 3.1~$\times$~4.1~\si{\milli\meter\squared} (Fig. \ref{fig:Implant_picture}) with a 0.076~mm, 0.5-oz copper trace on a 1.6 mm FR-4 printed circuit board. These parameters result in a simulated inductance of L = 12~nH at 2.4~GHz. To allow tuning the frequency between 2.2 and 2.6 GHz, the internal capacitor is adjustable between 310 fF and 440 fF. 

\section{Materials and Methods}

To validate the proposed infrared power and data delivery architecture, the neural response of degenerate rat retinas to electrical stimulation from a single electrode was measured with calcium imaging. Then, the implant power consumption budget is determined to evaluate the headroom for delivering complex stimulation patterns comprising multiple sequential pulses. 

\subsection{RGCs Response to Infrared-Powered Stimulation}

The response of retinal ganglion cells under infrared-powered stimulation is evaluated by generating spatial threshold maps of retinal ganglion cells around a single electrode. A map is realized for short pulse widths of 100~\si{\micro\second} and for longer pulses of 500~\si{\micro\second} to replicate a typical configuration used by the first generation of retinal implants~\cite{Dorn2013a}. The next subsections details how the spatial threshold maps are realized. 

\subsubsection{Implant test bench}
To deliver the stimulation pulses, the stimulator ASIC, photodiode, photovoltaic cell, crystal and passive components (C1, C2, C3, D1 from Figure \ref{fig:ASIC_architecture}) are assembled on a printed circuit board (Figure \ref{fig:methods2}). Then, the electrode driver pads are connected with cables to a 5~$\times$~5 electrode array assembled on a second printed circuit board. The power and data is sent to the implant using an 850~nm laser diode (L850P200, Thorlabs). The output power of the laser diode is adjusted by the laser driver (iC-NZ, iC-Haus Inc.) with a power meter to 35~mW. An ADRV9364-Z7020 System-on-Module controls the laser driver to encode the stimulation data with a binary amplitude shift keying (BASK) scheme. 

\subsubsection{Retina preparation}

Retina preparation is performed in accordance with the ethical protocol of the Animal Care and Ethics Committee of The University of Melbourne. Adult Royal College of Surgeons (RCS-p+) rats of either gender and older than 3 months are prepared. RCS rats have inherited retinal degeneration which causes their retina to lose most of its photoreceptors by 90 days after birth~\cite{Ray2010}.

The retina is injected with a fluorescent indicator dye through the optic nerve for calcium imaging. The dye is Oregon Green 488 BAPTA-1 solution (OGB- 1, Hexapotassium salt, Thermo Fisher Scientific, dissolved in deionised water). The retina preparation and calcium indicator loading is described in detail in a separate paper~\cite{Tong2019}.

The retina is mounted on the diamond electrode array with the ganglion cell layer facing up and held with a steel harp fitted with Lycra threads (SHD-25GH, Warner Instruments). The diamond array is assembled on a printed circuit board which constitutes the bottom of a 3D printed perfusion chamber. The chamber is perfused with a carbogenated Ames’ solution at a rate of 3-8~\si{\milli\liter\per\minute} held between 35\si{\degree}C and 37\si{\degree}C. The electrode array is kept around 2.5~mm away from the optic nerve.

Although the implant is designed to be placed epiretinally, the electrode array is placed subretinally in this demonstration to facilitate the experiment with calcium imaging. For maximum light transmission to an upright microscope, the retinal ganglion cells need to face the top of the microscope. Thus, the electrode array is placed on the bottom face (subretinally) in order to avoid obstructing the line of sight of the microscope. 

\subsubsection{Calcium imaging}

The retina preparation is imaged with a confocal microscope (Olympus FluoView FV1200) with a 10$\times$ and a 20$\times$ lens, for a field of view of either 318~$\times$~318~\si{\micro\meter\squared} or 633~$\times$~633~\si{\micro\meter\squared}. The calcium dye is excited with a 473~nm source, and images are captured at a rate of 7.8 Hz.

\subsubsection{Electrical stimulation}

\begin{figure}[h!t]
\centering
\includegraphics[width=0.5\textwidth]{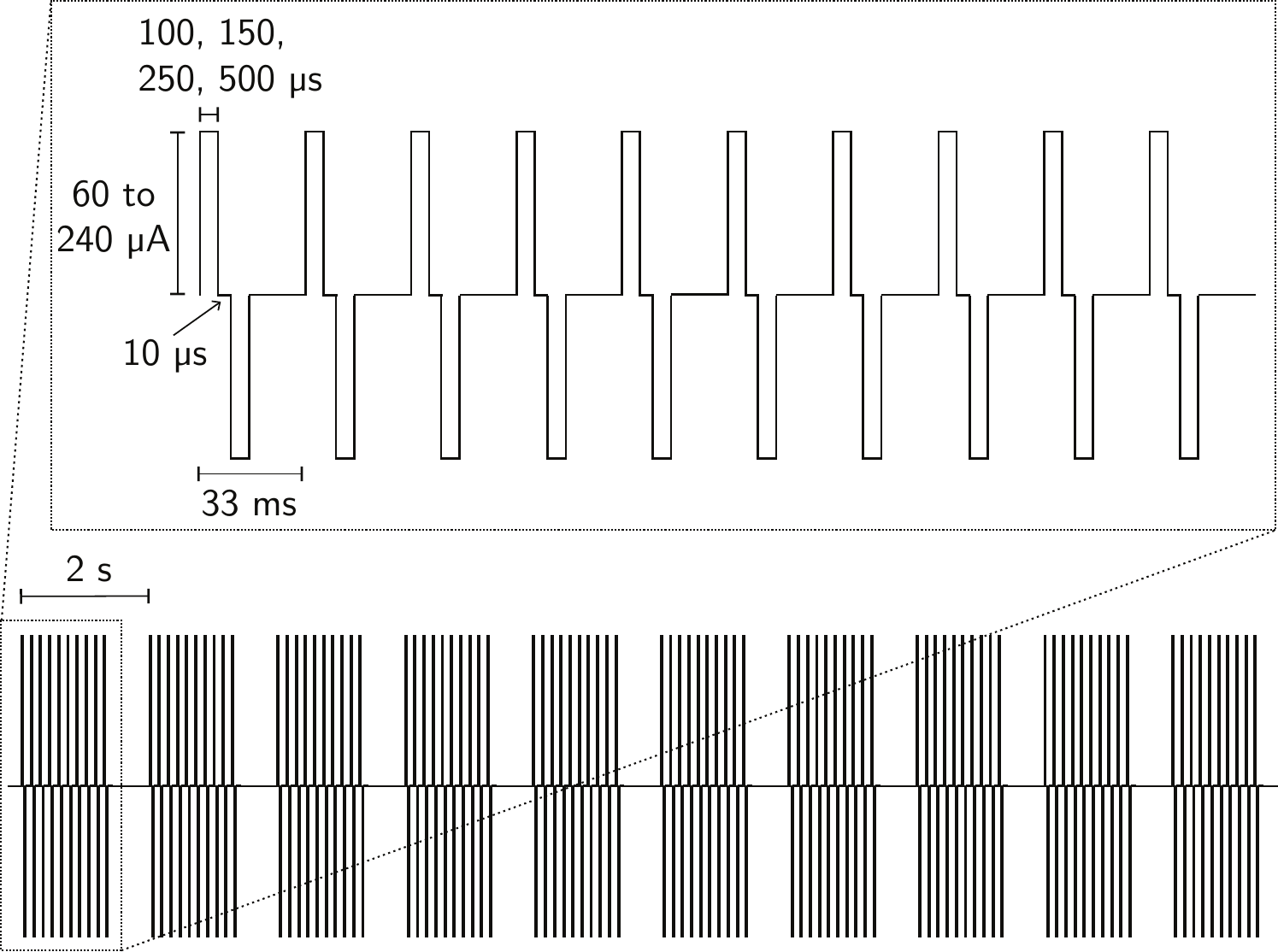}
\caption[Stimulation sequence for the experiments]{The stimulation sequence is composed of bursts of 10 pulses with a 33~ms period. The bursts are repeated 10 times with a 2~s period. This stimulation sequence is repeated for each combination of current (60~\si{\micro\ampere} to 240~\si{\micro\ampere} by steps of 20~\si{\micro\ampere}) and phase duration (100~\si{\micro\second}, 150~\si{\micro\second}, 250~\si{\micro\second} and 500~\si{\micro\second}).
}
\label{fig:stimulation_parameters}
\end{figure}

The electrical stimulation is delivered by the ASIC and consists of charge balanced, biphasic current-controlled pulses. The pulses are delivered with an anodic-first polarity, with phase durations of 100~\si{\micro\second} and 500~\si{\micro\second} with a 10~\si{\micro\second} interphase gap. The dynamic current copy architecture of the stimulation drivers requires a calibration phase prior to the stimulation whose duration is set to 30~\si{\micro\second}. The stimulation protocol is detailed in Figure \ref{fig:stimulation_parameters}. An Ag|AgCl wire acts as the return electrode and is placed in the perfusion chamber, 2~cm away from the stimulating electrodes. 

\subsubsection{Data analysis}
Electrical responses are evaluated by identifying rapid temporal changes in the fluorescence image. To achieve this, the response is evaluated by filtering the signal of each pixel with a temporal high-pass filter (with coefficients [2,1,-1,-2]), and then detecting activation by setting a threshold to the intensity within the area of each identified RGC at twice the standard deviation of the signal. The current threshold of each RGC is evaluated by fitting a sigmoid function to the neuron's response, and selecting the amplitude associated to a detected response in 50~\% of the cases. The data analysis is presented in detail in a separate paper~\cite{Tong2019}.

\subsection{Implant Power Budget}
The implant power budget is determined by first characterizing the photovoltaic cell to determine its power output. Then, the remaining power for stimulation is evaluated by subtracting the implant standby power consumption from the photovoltaic cell output power. Then, from the calcium imaging experiments, the required stimulation power is measured at the average stimulation threshold for a single electrode. From this measurement, the maximum achievable number of stimulation pulses per second (maximum repetition rate) can be determined given the available power. 

\subsubsection{Photovoltaic Cell Characterization}

The photovoltaic cell is characterized by tracing the current-voltage and power-voltage curves under illumination with a 35~mW laser beam collimated on the photosensitive surface. The curves are traced with a Keithley 4200A source measurement unit (SMU).  

\subsubsection{Available Stimulation Power}

The available stimulation power is derived from the implant power budget by subtracting the losses associated with the ocular medium, the photovoltaic cell and the implant standby power consumption from the 35~mW power source. The standby power consumption is measured via the voltage drop on a 10~\si{\ohm} shunt resistor after the photovoltaic cell. 

\subsubsection{Maximum Repetition Rate}

The maximum stimulation repetition rate is a key metric indicative of the capacity of the implant to eventually mimic neural code on a spike-by-spike basis~\cite{Shah2019}. This maximum rate is limited by the available power. To evaluate the maximum stimulation rate, the power consumption for a single electrode is measured while delivering a current at the average threshold required to elicit a response. The average thresholds are evaluated with calcium imaging for pulse widths of 100~\si{\micro\second}, 150~\si{\micro\second}, 250~\si{\micro\second}, 500~\si{\micro\second} with three different pieces of retina. Then, the maximum pulse rate that can be delivered on the array with the available power is estimated by dividing the available stimulation power by the power consumption for a single electrode. The result is then divided by the time slot duration (twice the pulse width plus a 10~\si{\micro\second} interphase gap and a 30~\si{\micro\second} calibration interval for balancing the currents of the anodic and cathodic phases). 

\section{Results}

\subsection{RGC Response to Infrared-Powered Stimulation}

\begin{figure}[H]
\centering
\includegraphics[width=0.5\textwidth]{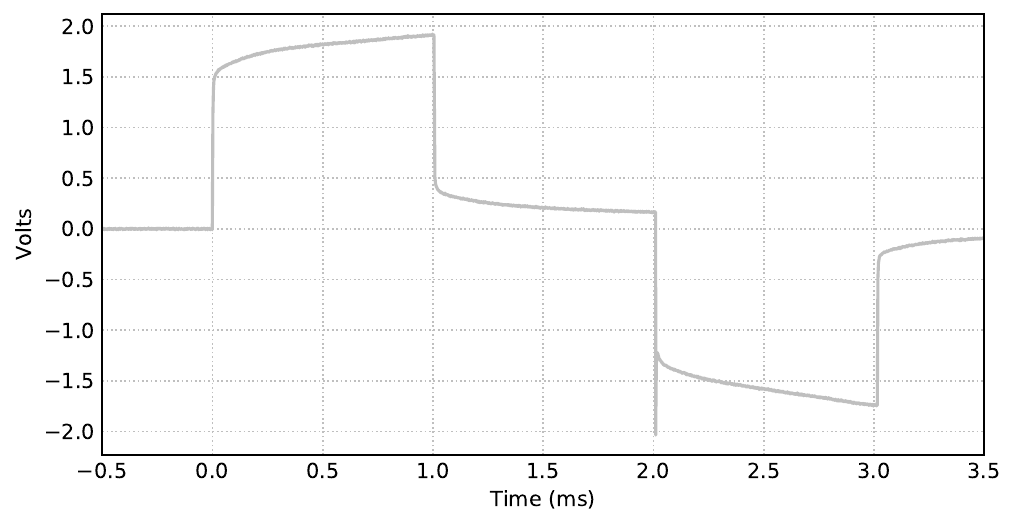}
\caption[Voltage waveform of a stimulation pulse]{Voltage waveform of a 1~ms stimulation pulse at 100~\si{\micro\ampere} in a physiological saline solution with an oscilloscope.}
\label{fig:fig_electrode_25_2}
\end{figure}

Firstly, the functionality of the device is verified by measuring the voltage waveform of a stimulation pulse with an oscilloscope (Figure \ref{fig:fig_electrode_25_2}).

\begin{figure}[H]
\centering
\includegraphics[width=0.5\textwidth]{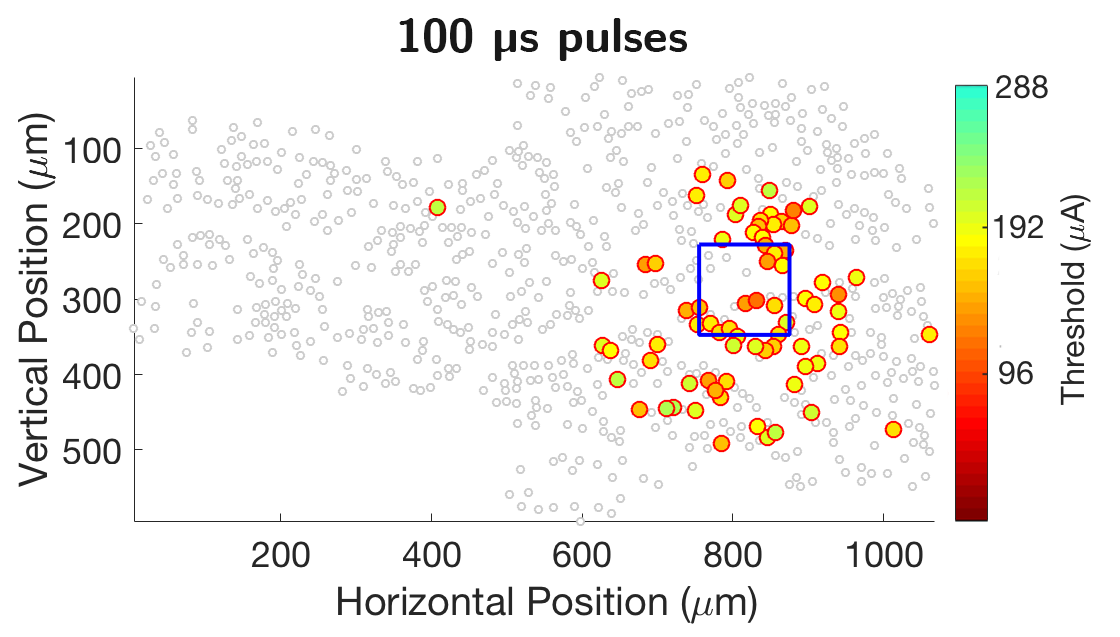}
\caption[Spatial threshold map for 100~\si{\micro\second}  pulses]{Spatial threshold map of retinal ganglion cells in a degenerate RCS rat retina for 100~\si{\micro\second} biphasic charge-balanced stimulation pulses.  }
\label{fig:wireless_thresholds_100us}
\end{figure}

\begin{figure}[H]
\centering
\includegraphics[width=0.5\textwidth]{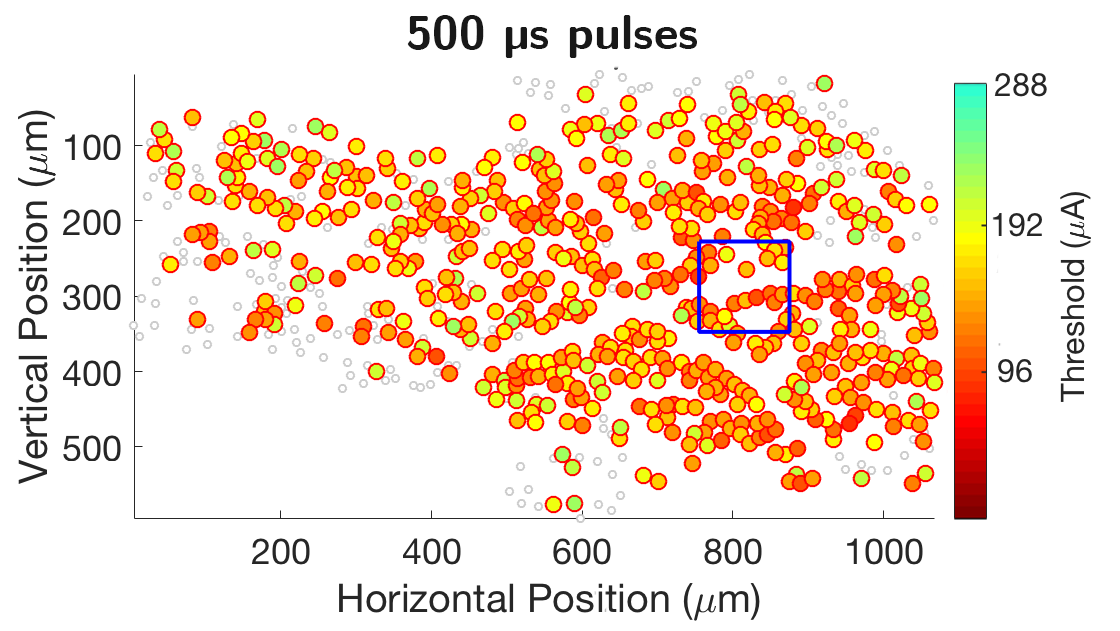}
\caption[Spatial threshold map for 500~\si{\micro\second} pulses]{Spatial threshold map of retinal ganglion cells in a degenerate RCS rat retina for 500~\si{\micro\second} biphasic charge-balanced stimulation pulses.}
\label{fig:wireless_thresholds_500us}
\end{figure}

Figures \ref{fig:wireless_thresholds_100us} and \ref{fig:wireless_thresholds_500us} present the RGC spatial threshold maps from 100~\si{\micro\second} and 500~\si{\micro\second} pulses with the implant being powered by a 35~mW laser. In the threshold map, each circle represents one RGC, with the color indicating the threshold current. The RGCs that couldn't be activated with the maximum available current are shown as open circles. The blue square indicates the electrode position. As reported previously, 100~\si{\micro\second} pulses lead to a more confined activation pattern. Using 500~\si{\micro\second} pulses, the larger activation spread is most likely due to the unintended stimulation of the axon bundles passing the electrode and network-mediated stimulation via bipolar cells~\cite{Tong2019, Chang2019}.

\subsection{Implant Power Budget}
\subsubsection{Photovoltaic Cell Characterization}

\begin{figure}[h!t]
\centering
\includegraphics[width=0.5\textwidth]{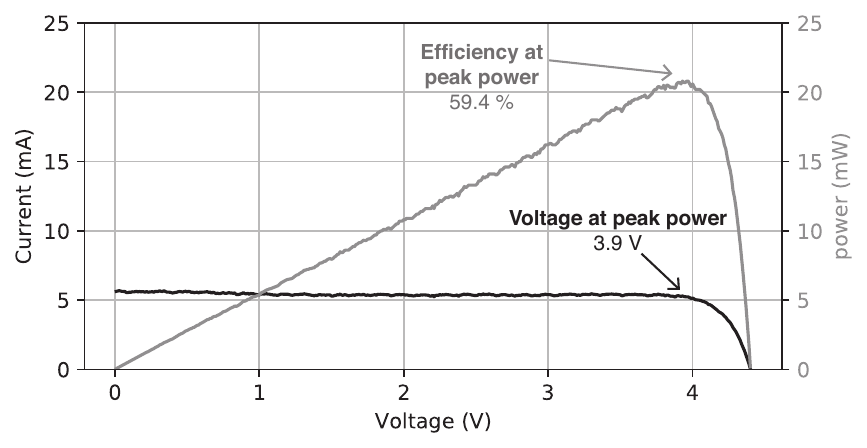}
\caption[Photovoltaic cell characterization]{Photovoltaic cell characterization at 850~nm with a 35 mW beam collimated within the sensitive area. The efficiency peaks at 59.4~\% at a voltage of 3.9~V.}
\label{fig:PV_cell}
\end{figure}

To evaluate the power budget of the implant, the photovoltaic cell is first characterized. Figure \ref{fig:PV_cell} presents its current-voltage and power-voltage curves with a 35~mW laser. At peak power, the cell outputs 3.9~V with an efficiency of 59.4~\%.

\subsubsection{Available Stimuation Power}

The implant power budget following the photovoltaic cell characterization is presented in Table \ref{table:power_consumption_procedure}.  With a maximum radiant power density of 4.06~\si{\milli\watt\per\milli\meter\squared}, a maximum of 36.5~mW can enter the eye for a 9~\si{\milli\meter\squared} photovoltaic cell, assuming uniform light distribution. The laser power entering the eye is set slightly below 35~mW. Because of the light absorption of the ocular medium, 20~\% of the light is absorbed (7.0~mW is dissipated), so that 28.0~mW reaches the photovoltaic cell~\cite{Boettner1962}. The latter then converts the beam reaching its surface with an efficiency of 59.4~\% (11.4~mW is dissipated). The ASIC consumes 3.5~mW of standby power consumption, which leaves 13.1~mW of power for delivering stimulation pulses.

\begin{table}[h!t]
\begin{center}
\caption{Implant Power Budget}
\label{table:power_consumption_procedure}
\begin{adjustbox}{width=0.5\textwidth}
\begin{tabular}{lc}
\hline
\textbf{Description  }                                                & \textbf{Power (mW)}\\
\hline
Laser                                                          & 35.0 \\
Eye optical losses (20~\% of 35~mW at 850~nm~\cite{Boettner1962}).   & -7.0 \\
PV cell power dissipation (59.4~\% of 28.0~mW)     & -11.4 \\
Implant standby power consumption                                    & -3.5 \\
\hline
Available stimulation power                                          & 13.1 \\
\hline
\end{tabular}
\end{adjustbox}
\end{center}
\end{table}

\subsubsection{Maximum Repetition Rate}

\begin{figure}[H]
\centering
\includegraphics[width=0.5\textwidth]{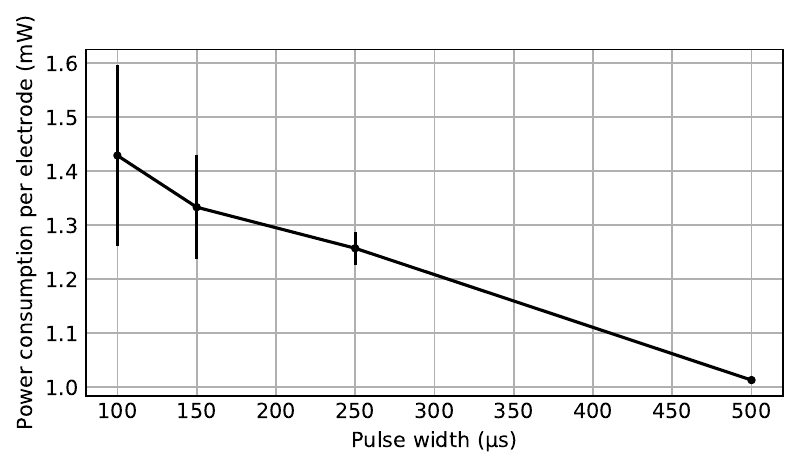}
\caption[Power consumption at the average stimulation threshold]{Power consumption of a single electrode at the average stimulation threshold for different pulse widths. The thresholds were averaged over three different retinas. The ASIC standby power consumption is excluded. }
\label{fig:power_consumption}
\end{figure}

During stimulation, the power consumption depends on the current amplitude required to trigger action potentials, which varies according to many factors, including electrode-neuron distance, electrode size, neuron physiology and pulse characteristics. For the current experiment conditions, the average thresholds for eliciting a response were calculated using calcium imaging for pulse widths of 100, 150, 250 and 500~\si{\micro\second}. Then, the maximum current drawn from the ASIC is measured during pulse delivery, and subtracted from the standby power consumption. This current is then multiplied by the PV cell voltage to obtain the power consumption of a single electrode at the average stimulation threshold, as shown in Figure \ref{fig:power_consumption}. 

\begin{figure}[h!t]
\centering
\includegraphics[width=0.5\textwidth]{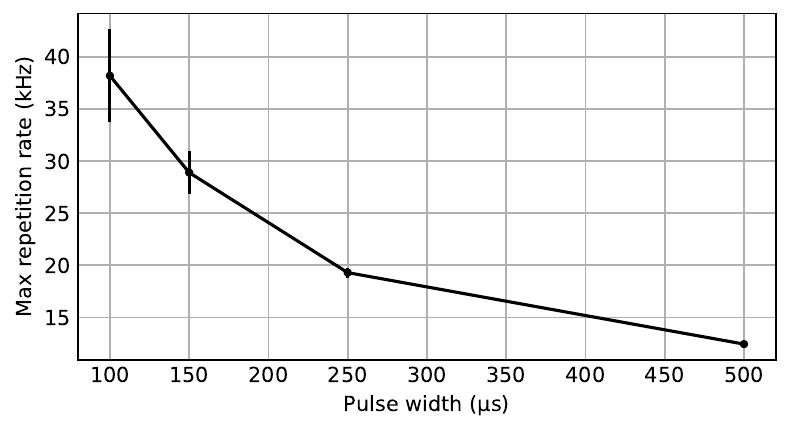}
\caption[Maximum stimulation pulse rate]{Maximum pulse rate that can be distributed on the electrode array based on the available power. It is calculated by dividing the available stimulation power (Table \ref{table:power_consumption_procedure}) by the power consumption per electrode (Figure \ref{fig:power_consumption}). The result is then multiplied by the number of time slots per second. A time slot is equal to twice the pulse width plus 10~\si{\micro\second} for the interphase gap and 30~\si{\micro\second} for the current copying calibration phase.}
\label{fig:repetition_rate}
\end{figure}

Figure \ref{fig:repetition_rate} presents the expected maximum stimulation rate that can be delivered on the array for sequential stimulation based on the measured stimulation thresholds and available power.

\section{Discussion}

The main objective of the paper is to evaluate the feasibility of infrared-powered retinal ganglion cells stimulation using a CMOS stimulator ASIC powered by a photovoltaic cell. The results from the previous section suggest it is possible within certain limitations. 

\subsection{Feasibility of Near-Infrared Power Delivery}

Figure \ref{fig:wireless_thresholds_100us} and \ref{fig:wireless_thresholds_500us} demonstrate that retinal ganglion cell response can be elicited using solely an infrared 35~mW laser beam as a power source. The 35~mW power source was chosen based on a photovoltaic cell with photosensitive dimensions of at least 3~$\times$~3~\si{\milli\meter\squared} in order to comply with the maximal permissible irradiance of 4~\si{\milli\watt\per\milli\meter\squared} at 850~nm. However, the prototype was realized with a commercially available 1.7~$\times$~1.7~\si{\milli\meter\squared} from Broadcom to reduce its development costs, as opposed to developing a cell with custom dimensions. Designing the implant with the appropriate photovoltaic cell dimensions is critical to achieve permissible irradiance levels. Moreover, further tests will need to be conducted to ensure safety of the device.

\subsection{Optimal pulse width considering the photovoltaic cell constraints}

In order to achieve wireless operation, a retinal prosthesis must use a stimulation strategy that optimizes the power consumption. Although the instantaneous power consumption is higher for shorter pulse widths (Figure \ref{fig:power_consumption}), the energy per pulse is lower (calculated by the multiplication of the power by the pulse duration). This is caused by the lower charge thresholds required to elicit a response with shorter pulse widths~\cite{Boinagrov2010a}. This effect is expected to plateau with pulse widths significantly below the cell chronaxie, at pulse widths around tens of microseconds~\cite{Merrill2005}. 

Practically, other factors limit the delivery of really short pulses. Shorter pulses require larger currents to deliver comparable amounts of charge. However, stimulators have a limited maximum current, especially in the case of wirelessly powered devices where high peak currents require a large energy reservoir. Additionally, the compliance voltage of the stimulator limits the pulse widths. At shorter pulse widths, the higher currents induce larger access voltages caused by the resistive component of the electrode-electrolyte impedance. With the proposed implant, the $\pm$2.7~V compliance limit prevented reliable elicitation of a response with pulses below 100~\si{\micro\second}.

The reported experiments have demonstrated effective stimulation with a single electrode and with a power significantly below the available power from the photovoltaic cell. This leaves headroom for activating multiple electrodes simultaneously. Figure \ref{fig:repetition_rate} presents the expected maximum repetition rate achievable given the experimental conditions. A higher repetition rate allows more accurate neural code reproduction in stimulation strategies based on a rapid sequence of electrical stimuli from a given dictionary of possibilities~\cite{Shah2019}. Rather than activating all electrodes simultaneously, as is common with subretinal implants, the implant optimally utilizes its power by sequentially activating electrodes with short, high-current pulses. %To prevent electrical crosstalk during concurrent stimulation, these electrodes should be separated by a minimum distance, and ideally with return electrodes~\cite{Flores2016}.

\subsection{Experiment Limitations}

In this experiment, the electrode array is located subretinally instead of epiretinally to preserve the line-of-sight between the RGCs and the confocal microscope objective lens. The electrodes are consequently separated from the RGCs by the thickness of the retina, which varies between 100~to~200~\si{\micro\meter}. Additionally, the 120~\si{\micro\meter} electrodes used in this experiment are relatively large compared to other experiments with electrode sizes as small as 5~\si{\micro\meter}~\cite{Hottowy2012,Sekirnjak2006a}. These two factors increase the stimulation thresholds substantially. Alternatively, transparent indium tin oxide electrodes could be placed epiretinally without obstructing the line of sight~\cite{Weitz2016}, but would not exactly reproduce the behavior of the diamond electrode array. With smaller 10~\si{\micro\meter} electrodes placed epiretinally, stimulating with biphasic electrical pulses of 0.05–0.1 ms result in thresholds in the order of 1~\si{\micro\ampere}~\cite{Sekirnjak2006a, Grosberg2017}. This requires close proximity of the electrodes to the ganglion cells, which is achievable in \textit{in-vitro} experiments, but can be highly challenging in a clinical context~\cite{Gregori2018, Ahuja2013}. Lower thresholds would reduce the power consumption and allow higher stimulation repetition rates. Moreover, using smaller electrodes is critical to attain the spatial resolution required for single-cell stimulation and to reduce the required energy per stimulation pulse.

\subsection{Comparison With Other Photovoltaic Architectures}

Other implant architectures previously demonstrated the capability to elicit responses using 0.2 to 10~\si{\milli\watt\per\milli\meter\squared} of irradiance at 905~nm~\cite{Mathieson2012}. These architectures target bipolar cells with subretinal electrodes. In contrast to these analog subretinal devices that use direct coupling between photodiodes and electrode drivers, the current study introduces a digital epiretinal architecture targeting retinal ganglion cells, powered and controlled by a digital infrared link. 

The NR600 photovoltaic system also demonstrates digital photovoltaic stimulation. However, it differs by its internal camera and stimulation strategy, where electrodes are activated in parallel with preconfigured patterns, reportedly targeting bipolar cells~\cite{yanovitch_new_2022, borda_advances_2022}. In contrast, this paper combines an external camera with a faster data link, allowing sequential electrode activation without a predetermined pattern to target retinal ganglion cells. Although multiple challenges remain in achieving precise single cell activation using this architecture, sequential stimulation is crucial for future epiretinal implants aiming to reproduce the retinal code on a spike-by-spike basis.

Different neural types respond very differently to electrical stimulation. Bipolar cells respond preferentially to longer pulse widths with low currents (around 25 ms) and retinal ganglion cells respond preferentially to shorter pulse widths with higher currents (around 0.1 ms)~\cite{Tong2019, Weitz2016, Freeman2010}. This leads to very different requirements in terms of stimulation strategies. When targeting bipolar cells, the longer pulse widths impose parallel stimulation strategies where most electrodes are activated simultaneously to achieve a reasonable refresh rate. Architectures based on photodiode arrays are well tailored to this approach, as each photodiode transduces the energy to the electrode to which they are coupled.

For retinal ganglion cells, shorter pulse widths of around 0.1~ms allow for multiple time windows in which to deliver stimulation pulses within the image integration time of the brain~\cite{Chichilnisky2003}. Thus, electrodes could be stimulated sequentially, one at a time or in small groups. In terms of power delivery, this corresponds to concentrating the available power to the few simultaneously active electrodes. The photovoltaic cell approach proposed in this paper has the capacity of concentrating the total incident optical power on the active electrode, thus allowing shorter pulses at higher currents, as deemed preferable for RCG stimulation.

\section{Conclusion}

We presented an implant architecture based on an optical power and data link capable of eliciting a response in retinal ganglion cells while retaining the flexibility of a digital stimulation controller. The limited permissible radiant power entering the eye is sufficient to power the digital stimulation ASIC, ancillary circuits and deliver stimulation pulses that elicit a response in retinal ganglion cells. The proposed solution promises higher safety and reliability due to the possibility of encapsulating the device in a hermetic package without wires protruding of the implant and through the eyeball. With the goal of achieving meaningful visual acuity gains, next generations of epiretinal prostheses will need to deliver stimulation pulses that reproduce the neural code at a spatial resolution of cellular scale. Towards that goal, one of the next major challenges will be the realisation of a closed-loop device capable of wirelessly stimulating and recording with high electrode density. 

\section{Acknowledgements}
Authors gratefully acknowledge insightful discussions with Rob Hilkes, Tommy Rossignol, Émile Laplante, Patrice Buteau, Anne Bruneau and Jean Wilson.

\section{Conflicts of Interest}
SP was a shareholder in iBIONICS, a company developing a diamond based retinal implant. SP and DG are shareholders and directors of Carbon Cybernetics, a company developing brain-machine neural interfaces.

\bibliographystyle{IEEEtran}

\bibliography{References/library}

\end{document}